# Peculiarities of low-frequency vibrational dynamics and low-temperature heat capacity of double-walled carbon nanotubes

M. V. Avramenko, S. B. Rochal

*Southern Federal University, Rostov-on-Don, Russia*
E-mail: avramenko.marina@gmail.com

A continuous model of double-walled carbon nanotube (DWCNT) low-frequency dynamics was constructed. In the frame of the approach proposed formation of a DWCNT low-frequency phonon spectrum from the ones of corresponding single-walled carbon nanotubes (SWCNTs) was considered. Environmental influence on the individual DWCNT phonon spectrum was studied. A combined method of van der Waals interlayer coupling coefficients evaluation was proposed, and it is based on Raman spectroscopy data and known values of graphite elastic moduli. Also DWCNT low-temperature specific heat was calculated. In the model applicability region (at temperatures lower than 35 K) DWCNT specific heat turned out to be significantly lower than the specific heat sum of corresponding individual single-walled SWCNTs. This effect is caused mainly by the interlayer coupling instead of DWCNT interaction with environment.

## 1. Введение

Двустенные углеродные нанотрубки (ДУНТ) можно рассматривать как самые маленькие в мире коаксиальные кабели, что в сочетании с их уникальными электронными свойствами [1, 2] обусловливает будущее применение ДУНТ в наноэлектронике. И, поскольку с миниатюризацией электронных устройств до наноразмеров их производительность и стабильность во многом зависит от эффективности охлаждения, исследование тепловых свойств ДУНТ представляет особый интерес. В частности, теплоемкость ДУНТ является одной из тех важных тепловых характеристик, которые на сегодняшний день остаются крайне малоизученными. Нам известно всего три работы, представляющие экспериментальные зависимости теплоемкости ДУНТ от температуры: одна [3] в широком температурном диапазоне $T$=223 – 573 K, две другие [4, 5] при сверхнизких температурах до 20 и 30 K соответственно. Теоретически теплоемкость ДУНТ также мало изучена: авторы [6] исследовали теплоемкость соразмерных ДУНТ, представляя слои трубки как связанные осцилляторы, а в работе [7] был использован метод структурной молекулярной динамики. При этом авторы [6] утверждают, что низкотемпературная теплоемкость ДУНТ оказывается ниже, чем суммарная теплоемкость их отдельных слоев (примерно в 1.5 раза при $T$=10 K). В работе [7], напротив, авторы приходят к заключению, что теплоемкость углеродных нанотрубок в пренебрежимо малой степени зависит как от диаметра трубок, так и количества образующих их слоев. Заметим, что в эксперименте [5] наблюдалось подтверждение выводов, сделанных авторами работы [6].

Таким образом, современные представления о теплоемкости ДУНТ являются неполными и, в некоторой степени, противоречивыми. Для исследования низкотемпературной теплоемкости ДУНТ мы развиваем недавно предложенную континуальную модель [8] низкочастотной динамики трубчатых наноструктур. Модель [8] основана на тех же идеях, что и более ранняя работа [9]. В рамках подходов [8, 9] графен и свернутые из него индивидуальные нанотрубки рассматриваются как двумерные (не имеющие макроскопической толщины) мембраны, характеризуемые всего тремя упругими коэффициентами. Работы [8, 9] описывают только низкочастотные фононные моды нанотрубок, происходящие из акустических мод графена. При учете ван-дер-Ваальсова взаимодействия между двойными стенками континуальная модель позволяет описать и низкочастотную динамику ДУНТ. В рамках континуального подхода в [8] была определена связь между радиальными дыхательными модами ОУНТ и дыхательно-подобными модами ДУНТ. В работе [10] данный подход был модифицирован для учета взаимодействия между нанотрубкой и ее окружением (упругой средой или другими нанотрубками в пучке). Оказалось, что в результате данного взаимодействия низкочастотный фононный спектр ОУНТ существенным образом перестраивается, а теплоемкость ОУНТ при сверхнизких температурах падает более чем на порядок, что соответствует доступным на сегодняшний день экспериментальным данным.

Целью настоящей работы является развитие модели низкочастотной динамики ДУНТ. В отличие от работы [8], мы учитываем не только радиальное, но и тангенциальное взаимодействие между стенками нанотрубки, что позволяет нам более подробно проанализировать связь между низкочастотным спектром ДУНТ и спектрами двух составляющих ее индивидуальных нанотрубок. Также, следуя логике [10], мы рассматриваем перестройку спектра индивидуальной ДУНТ под влиянием окружения. Полученная модель применяется для расчета низкотемпературной и сверхнизкотемпературной теплоемкости ДУНТ.

Работа структурирована следующим образом. В разделе 2 представлена модель низкочастотной динамики ДУНТ, взаимодействующей с окружением, и обсуждаются ее ограничения. В разделе 3 рассматривается образование низкочастотного фононного спектра ДУНТ из фононных спектров составляющих ее ОУНТ, а также оцениваются коэффициенты межслоевого взаимодействия в ДУНТ через модули упругости графита. В разделе 4 развитый нами подход применяется для расчета теплоемкости ДУНТ, а в Заключении приводятся обсуждение результатов и выводы.

## 2. Модель низкочастотной динамики ДУНТ, взаимодействующей с окружением

Следуя континуальной модели [8], выражение для плотности свободной энергии двумерной мембраны можно записать в виде:

$$g = \frac{\lambda}{2}\varepsilon_{ii}^2 + \mu\varepsilon_{ij}^2 + 2K(\Delta H)^2, \qquad (1)$$

где $\lambda$ и $\mu$ – плоские аналоги коэффициентов Ламе, $K$ – прогибная упругость, $\Delta H = H - H_0$, $H_0$ – равновесная средняя кривизна мембраны, $H$ – кривизна, приобретаемая мембраной в результате ее деформации, $\varepsilon_{ij}$ – двумерный тензор деформации, зависящий в общем случае от $H_0$ и трехмерного поля смещений мембраны $\mathbf{u} = (u_r, u_\varphi, u_z)$. Последнее в цилиндрической системе координат зависит от угла $\varphi$ и переменной $z$, равной расстоянию вдоль оси недеформированной мембраны. Изменение кривизны, линеаризованное по полю смещений и его производным, имеет вид:

$$\Delta H = -(\Delta_s u_r)/(2R^2), \qquad (2)$$

где $\Delta_s = 1 + \partial_\varphi^2 + R^2 \partial_z^2$.

Заметим, что связь (в форме последнего слагаемого (1)) между изменением средней кривизны мембраны и энергией ее изгиба, была установлена еще в пионерских работах Канхама и Хелфриша [11, 12]. Также отметим, что плотность энергии (1) тождественна аналогичным выражениям из работ [8, 10] и с точностью до определения упругих модулей и выбора способа параметризации поля смещений эквивалентна [9]. Однако, как показывают результаты [13], энергия изгиба цилиндрической мембраны, в принципе, может задаваться и гораздо более сложным выражением, куда дополнительно к $\Delta_s$ входят еще два дифференциальных инварианта, вклады которых в развиваемой простейшей модели мы учитывать не будем.

При рассмотрении низкочастотной динамики ДУНТ нам необходимо учесть ван-дер-Ваальсово взаимодействие между ее слоями, основной вклад которого в полную энергию системы дают три компоненты:

$$U_r = \frac{G_r}{2}\int \left(u_r^{(1)} - u_r^{(2)}\right)^2 d\varphi dz, \quad U_\varphi = \frac{G_\varphi}{2}\int \left(u_\varphi^{(1)} - u_\varphi^{(2)}\right)^2 d\varphi dz, \quad U_z = \frac{G_z}{2}\int \left(u_z^{(1)} - u_z^{(2)}\right)^2 d\varphi dz, \qquad (3)$$

где $U_r$, $U_\varphi$ и $U_z$ – вклады радиального, тангенциальных поперечного и продольного взаимодействий, величины $G_r$, $G_\varphi$ и $G_z$ – материальные константы, описывающие данные взаимодействия, а $u_j^{(1)}$ и $u_j^{(2)}$, где $j=r$, $\varphi$, $z$ – соответствующие компоненты полей смещений внутренней и внешней нанотрубок, образующих ДУНТ. Мы также предполагаем, что внешняя нанотрубка взаимодействует с окружением упругим образом, при этом, следуя [10], мы считаем, что окружение просто пытается скомпенсировать радиальное движение внешней нанотрубки. Аналогичные члены возможны и для тангенциального движения, однако они должны быть существенно меньше [10], и для упрощения модели не учитываются.

Уравнения движения мембраны, решения которых определяют зависимости поля смещений $\mathbf{u}$ от координат и времени, находятся путем вариации функционала:

$$A = \int \left[ R_1 g(\mathbf{u}^{(1)}) + R_2 g(\mathbf{u}^{(2)}) - \frac{\rho}{2}\left(R_1 (\dot{\mathbf{u}}^{(1)})^2 + R_2 (\dot{\mathbf{u}}^{(2)})^2\right) + U_r + U_\varphi + U_z + \frac{C}{2}R_2 \left(u_r^{(2)}\right)^2 \right] d\varphi dz dt \qquad (4)$$

где $t$ – время, $R_1$ и $R_2$ – радиусы внутренней и внешней нанотрубок соответственно, $\rho$ – одинаковая для обеих мембран поверхностная плотность и $C$ – пиннинг-коэффициент, описывающий радиальное взаимодействие внешней нанотрубки с окружением (упругой средой или другими нанотрубками в пучке) [10]. Получившиеся уравнения имеют следующий вид:

$$\ddot{u}_r^{(1)}\rho R_1 = -(\lambda+2\mu)\left(\frac{u_r^{(1)}}{R_1}+\frac{\partial u_\varphi^{(1)}}{R_1\partial\varphi}\right)-\lambda\frac{\partial u_z^{(1)}}{\partial z}-K\frac{1}{R_1^3}\Delta_s^2 u_r^{(1)}+G_r\left(u_r^{(2)}-u_r^{(1)}\right),$$

$$\ddot{u}_\varphi^{(1)}\rho R_1 = \frac{(\lambda+2\mu)}{R_1}\left(\frac{\partial u_r^{(1)}}{\partial\varphi}+\frac{\partial^2 u_\varphi^{(1)}}{\partial\varphi^2}\right)+(\lambda+\mu)\frac{\partial^2 u_z^{(1)}}{\partial\varphi\partial z}+\mu R_1\frac{\partial^2 u_\varphi^{(1)}}{\partial z^2}+G_\varphi\left(u_\varphi^{(2)}-u_\varphi^{(1)}\right),$$

$$\ddot{u}_z^{(1)}\rho R_1 = (\lambda+\mu)\frac{\partial^2 u_\varphi^{(1)}}{\partial\varphi\partial z}+(\lambda+2\mu)R_1\frac{\partial^2 u_z^{(1)}}{\partial z^2}+\lambda\frac{\partial u_r^{(1)}}{\partial z}+\mu\frac{\partial^2 u_z^{(1)}}{R_1\partial\varphi^2}+G_z\left(u_z^{(2)}-u_z^{(1)}\right),$$

$$\ddot{u}_r^{(2)}\rho R_2 = -CR_2 u_r^{(2)}-(\lambda+2\mu)\left(\frac{u_r^{(2)}}{R_2}+\frac{\partial u_\varphi^{(2)}}{R_2\partial\varphi}\right)-\lambda\frac{\partial u_z^{(2)}}{\partial z}-K\frac{1}{R_2^3}\Delta_s^2 u_r^{(2)}+G_r\left(u_r^{(1)}-u_r^{(2)}\right), \qquad (5)$$

$$\ddot{u}_\varphi^{(2)}\rho R_2 = \frac{(\lambda+2\mu)}{R_2}\left(\frac{\partial u_r^{(2)}}{\partial\varphi}+\frac{\partial^2 u_\varphi^{(2)}}{\partial\varphi^2}\right)+(\lambda+\mu)\frac{\partial^2 u_z^{(2)}}{\partial\varphi\partial z}+\mu R_2\frac{\partial^2 u_\varphi^{(2)}}{\partial z^2}+G_\varphi\left(u_\varphi^{(1)}-u_\varphi^{(2)}\right),$$

$$\ddot{u}_z^{(2)}\rho R_2 = (\lambda+\mu)\frac{\partial^2 u_\varphi^{(2)}}{\partial\varphi\partial z}+(\lambda+2\mu)R_2\frac{\partial^2 u_z^{(2)}}{\partial z^2}+\lambda\frac{\partial u_r^{(2)}}{\partial z}+\mu\frac{\partial^2 u_z^{(2)}}{R_2\partial\varphi^2}+G_z\left(u_z^{(1)}-u_z^{(2)}\right).$$

Заметим, что, следуя логике [10], упругие пиннинг-члены, ответственные за возвращающие силы, соответствующие тангенциальному и вращательному движению внешней нанотрубки как целого, должны быть существенно, по крайней мере, на порядок, меньше члена, ответственного за радиальную возвращающую силу, и поэтому не содержатся ни в функционале (4), ни в уравнениях движения (5).

Чтобы решить систему (5), подставим $u_j = u_j^0 \exp(i(kz+n\varphi-\omega t))$, где $j=r, \varphi, z$, $n$ – целое волновое число, $k$ – модуль однокомпонентного волнового вектора, $\omega$ – циклическая частота. Обозначим: $X = R^2 k^2 + n^2 - 1$, $\psi = (\lambda+\mu)nk$, $\alpha(R) = (\lambda+2\mu)/R + KX^2/R^3 - R\rho\omega^2$, $\beta(R) = (\lambda+2\mu)n^2/R + \mu k^2 R - R\rho\omega^2$, $\gamma(R) = (\lambda+2\mu)k^2 R + \mu n^2/R - R\rho\omega^2$, $\sigma(R) = i(\lambda+2\mu)n/R$, $\tau = ik\lambda$. Тогда равенство нулю определителя динамической матрицы **M** задает шесть действительных законов дисперсии $\omega_j = \omega_j(k,n)$:

$$\mathbf{M} = \begin{bmatrix} \alpha(R_1)+G_r & \sigma(R_1) & \tau & -G_r & 0 & 0 \\ -\sigma(R_1) & \beta(R_1)+G_\varphi & \psi & 0 & -G_\varphi & 0 \\ -\tau & \psi & \gamma(R_1)+G_z & 0 & 0 & -G_z \\ -G_r & 0 & 0 & \alpha(R_2)+G_r+CR_2 & \sigma(R_2) & \tau \\ 0 & -G_\varphi & 0 & -\sigma(R_2) & \beta(R_2)+G_\varphi & \psi \\ 0 & 0 & -G_z & -\tau & \psi & \gamma(R_2)+G_z \end{bmatrix} \qquad (6)$$

Для некоторых высокосимметричных мод матрица (6) существенно упрощается, поскольку становится квазидиагональной. Так, например, низкочастотный спектр ДУНТ содержит всего четыре моды с $k=0$ и $n=0$. Квазидиагональность матрицы **M** позволяет легко определить частоты этих мод. Первая пара их них – так называемые дыхательно-подобные моды (ДПМ) [14, 15] ДУНТ. Их частоты можно найти как решения следующего квадратного уравнения:

$$\left[R_1\left(\omega_1^2-\omega^2\right)+G_r'\right]\left[R_2\left(\omega_2^2-\omega^2\right)+G_r'+CR_2/\rho\right]-G_r'^2 = 0, \qquad (7)$$

где $G_r' = G_r/\rho$ и $\omega_i^2 = (\lambda+2\mu)/\rho R_i^2$. Заметим, что для индивидуальной нанотрубки, не взаимодействующей с окружением, величина $(\lambda+2\mu)/\rho R^2$ в точности равна квадрату частоты ее радиально-дыхательной моды (РДМ). Именно этот факт позволил определить зависимость $G_r$ от диаметров нанотрубок и прояснить связь между частотами ДПМ и РДМ в работе [8].

Вторая пара мод ДУНТ с $k=0$ и $n=0$ характеризуется частотами:

$$\omega_i = \sqrt{\frac{G_i(R_1+R_2)}{\rho R_1 R_2}}, \qquad (8)$$

где $i=\varphi, z$. Следует отметить, что первая из этих мод соответствует относительному вращению нанотрубок, а вторая – их относительному проскальзыванию. Естественно, без учета тангенциального взаимодействия между слоями нанотрубки частоты обеих мод становятся в точности равными нулю. Поэтому учет тангенциального взаимодействия в любой динамической модели, предлагаемой для расчета теплоемкости многослойных трубок, является обязательным. Именно на величины (8) в точке $k=0$ приподнимаются соответствующие дисперсионные ветки ДУНТ при взаимодействии звуковых дисперсионных веток индивидуальных ОУНТ. Данный вопрос мы рассмотрим подробнее в следующем разделе, а в заключение этого кратко обсудим область применимости предлагаемой модели.

Любая континуальная модель может быть применима, если длина волны рассматриваемых волн существенно больше межатомных расстояний. Так как волновое число $n$ равно числу осцилляций, которые делает рассматриваемая волна при обходе по периметру трубчатой мембраны, то фононы в ней можно характеризовать длиной эффективного волнового вектора

$$k_{eff} = \sqrt{k^2 + \left(\frac{n}{R}\right)^2}, \qquad (9)$$

где под $R$ следует понимать внутренний (меньший) радиус. Длина (9), в свою очередь, определяет эффективную длину волн $2\pi/k_{eff}$, которая должна быть в несколько раз больше межатомного расстояний в структуре, что и является главным ограничением континуальной модели. Учтя, что период решетки графена $a=0.246$ nm, мы считаем континуальную модель применимой в случае $k_{eff} < k_{lim}$, где $k_{lim} \sim 7$ nm$^{-1}$.

Заметим, что, в отличие от подходов типа [6, 7], явно учитывающих периодичность ДУНТ, континуальная модель применима и для несоразмерных нанотрубок. Как известно, ДУНТ обладает периодичностью, только если отношение периодов образующих ее ОУНТ равно рациональному числу. Тем не менее, значительное число структурно идентифицированных ДУНТ являются несоразмерными. На наш взгляд, это никак не должно помешать существованию в подобных ДУНТ фононных дисперсионных веток, происходящих из длинноволновых акустических мод графена. В подтверждение нашего утверждения напомним, что при малых волновых векторах акустические ветки несоразмерных 3D структур вполне четко идентифицируются в экспериментах по неупругому рассеянию нейтронов.

Несоразмерность приводит к тому, что с ростом длины волнового вектора данные ветки расширяются и исчезают. Таким образом, звук с малыми волновыми векторами в таких структурах вполне может рассматриваться в континуальном приближении. Аналогично, мы полагаем, что для расчета низкочастотного спектра ДУНТ континуальная модель также вполне приемлема. В дополнение заметим, что если размер первой зоны Бриллюэна соразмерной ДУНТ короче вектора $k$, характеризующего данный фонон в континуальном приближении, то, зная зонную структуру данной ДУНТ, всегда можно найти эквивалентный волновой вектор в ее первой зоне Бриллюэна. Разумеется, континуальная модель не может учесть возможные слабые искажения фононных веток, появляющиеся на границах сложенной первой зоны.

### 3. Влияние на низкочастотную динамику ДУНТ взаимодействий между образующими ее нанотрубками и взаимодействия с окружением

Для того чтобы практически применять предложенную модель, нам прежде всего требуется оценить значения используемых в ней материальных констант. Следуя [8], будем использовать систему единиц, в которой размеры НТ задаются в nm, длина волнового вектора $k$ – в nm$^{-1}$, а решения det|**M**|=0 соответствуют частотам фононных мод в cm$^{-1}$. Для характеристики упругих свойств графенового листа возьмем те же оценки, что и в [8]: $\lambda/\rho \approx 2400$ cm$^{-2}$ nm$^2$, $\mu/\rho \approx 5200$ cm$^{-2}$nm$^2$, $K/\rho \approx 12.5$ cm$^{-2}$ nm$^4$. Величина модуля $C$ зависит от материала, в котором находится ДУНТ. Рассмотрим вначале индивидуальную ДУНТ, не взаимодействующую с окружением ($C=0$), и обсудим связь ее низкочастотного спектра со спектрами двух образующих ее ОУНТ.

Величину константы радиального межслоевого взаимодействия $G_r$ для индивидуальной ДУНТ проще и надежней всего оценить на основе уравнения (7), связывающего частоты ее ДПМ с частотами РДМ индивидуальных ОУНТ. К сожалению, моды (8) в ДУНТ не являются резонансными, и при помощи современных спектроскопических методов экспериментально наблюдаться не могут. Поэтому оценить константы $G_\varphi$ и $G_z$ можно лишь косвенно. Так как расстояние $h$ между слоями ДУНТ, как правило, достаточно близко по значению к межплоскостному расстоянию в графите [16], можно оценить константы взаимодействий между слоями ДУНТ приближенно, выразив их через упругие модули графита.

Как известно, упругая энергия, отнесенная к единице объема кристалла, характеризуемого тензором деформаций $\varepsilon_i$, равна $c_{ij}\varepsilon_i\varepsilon_j/2$, где $c_{ij}$ – матрица констант упругой жесткости кристалла [17]. С другой стороны, энергия межслоевого взаимодействия в ДУНТ задается формулами (3). Вводя следующие обозначения: $l$ – длина ДУНТ, $V$ – ее средний межслоевой объем, численно равный $\pi l h(R_1 + R_2)$ и выражая компоненты тензора деформаций $\varepsilon_i$ в виде $\varepsilon_3 = (u_r^{(1)} - u_r^{(2)})/h$, $\varepsilon_4 = (u_\varphi^{(1)} - u_\varphi^{(2)})/h$, $\varepsilon_5 = (u_z^{(1)} - u_z^{(2)})/h$, получаем

$$\frac{\pi l}{2} c_{33} \frac{R_1 + R_2}{R_1 - R_2} \left(u_r^{(1)} - u_r^{(2)}\right)^2 \approx G_r \pi l \left(u_r^{(1)} - u_r^{(2)}\right)^2,$$

$$\frac{c_{44}}{2} \pi l \frac{R_1 + R_2}{R_1 - R_2} \left(u_\varphi^{(1)} - u_\varphi^{(2)}\right)^2 \approx G_\varphi \pi l \left(u_\varphi^{(1)} - u_\varphi^{(2)}\right)^2, \qquad (10)$$

$$\frac{c_{44}}{2} \pi l \frac{R_1 + R_2}{R_1 - R_2} \left(u_z^{(1)} - u_z^{(2)}\right)^2 \approx G_z \pi l \left(u_z^{(1)} - u_z^{(2)}\right)^2.$$

Из соотношений (10) следует, что

$$G_r \approx \frac{c_{33}}{2} \frac{R_1 + R_2}{R_1 - R_2}, \quad G_\varphi \approx G_z \approx \frac{c_{44}}{2} \frac{R_1 + R_2}{R_1 - R_2}. \quad (11)$$

Таким образом, для констант межслоевого взаимодействия в ДУНТ приближенно выполняются следующие соотношения:

$$\frac{G_r}{G_\varphi} \approx \frac{G_r}{G_z} \approx \frac{c_{33}}{c_{44}} \approx \left(\frac{v_{\text{LA}}}{v_{\text{TA}}}\right)^2 \approx 7.75, \qquad (12)$$

где $v_{\text{LA}}$ и $v_{\text{TA}}$ − скорости распространения продольной и поперечной акустических волн в графите. Согласно литературным данным [18], при атмосферном давлении $c_{33}$~38.7 GPa, $c_{44}$~5 GPa, $v_{\text{LA}}$~4.14 и $v_{\text{TA}}$~1.48 km/s, что дает отношение между $G_r$ и $G_\varphi$ или $G_z$, приблизительно равное 7.75 раз. Заметим, что симметрия ДУНТ предполагает различие величины констант $G_\varphi$ и $G_z$, однако численно оценить разницу между ними пока не представляется возможным, поэтому далее мы будем полагать их равными друг другу. Также заметим, что формула (11) позволяет получить лишь грубую оценку $G_r$, поскольку приводит к ошибкам в десятки процентов по сравнению с оценкой на основе данных спектроскопии комбинационного рассеяния света (КРС) и уравнения (7). Поэтому в дальнейшем мы будем брать спектроскопическую оценку $G_r$, а для получения $G_\varphi$ и $G_z$ использовать формулу (12).

Рис. 1 демонстрирует связь между низкочастотными спектрами свободных ОУНТ (22, 14) и (40, 1) и образованной из них ДУНТ (22, 14)@(40, 1). Значения коэффициентов межслоевого взаимодействия в данной ДУНТ были определены на основе экспериментальных данных [19]: $G_r/\rho$~3844 cm$^{-2}$nm, $G_\varphi/\rho = G_z/\rho$~496 cm$^{-2}$nm. Прежде всего отметим, что в низкочастотном спектре ДУНТ можно найти целый ряд дисперсионных веток, довольно хорошо совпадающих с ветками в спектрах исходных одностенных нанотрубок. Например, в ДУНТ скорости распространения поперечной вращательной ($v_{\text{TA}}$) и продольной ($v_{\text{LA}}$) акустической волн, имеющих линейную дисперсию при $k \to 0$, совпадают с таковыми в ОУНТ:

$$v_{\text{LA}} = \sqrt{\frac{4\mu(\lambda + \mu)}{\rho(\lambda + 2\mu)}}, \quad v_{\text{TA}} = \sqrt{\frac{\mu}{\rho}}. \qquad (13)$$

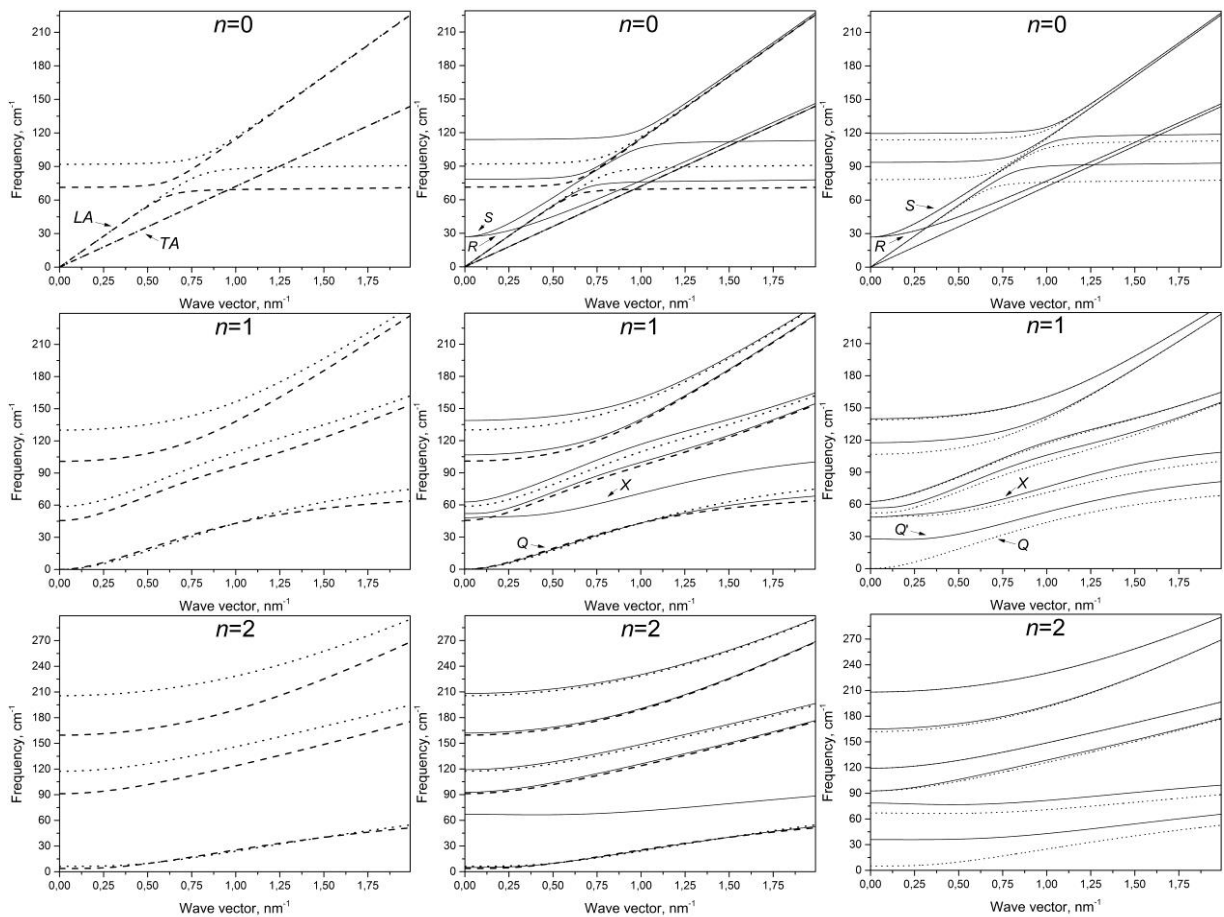

Рис. 1. Связь между низкочастотными фононными спектрами свободных ОУНТ (22, 14) и (40, 1) и ДУНТ (22, 14)@(40, 1). Спектры ОУНТ (22, 14) и (40, 1) показаны в первом столбце штриховыми и пунктирными линиями соответственно. Во втором столбце сверху дополнительно наложен фононный спектр свободной ДУНТ, показанный сплошными линиями. Наконец, в последнем столбце тот же спектр свободной ДУНТ перерисован пунктирной линией, а сверху сплошной линией изображен спектр ДУНТ, взаимодействующей с окружением. Графики, представленные в каждой строке, соответствуют дисперсионным кривым фононного спектра с одинаковым значением волнового числа $n$.

Тем не менее, низкочастотный фононный спектр ДУНТ существенно отличается от простой суперпозиции фононных спектров составляющих ее ОУНТ. Так происходит за счет ван-дер-Ваальсовых взаимодействий между слоями ДУНТ. Заметим, что и у свободной ОУНТ, и у свободной ДУНТ имеется ровно четыре голдстоуновских степени свободы (ГСС): вращение и движение как целого вдоль оси нанотрубки (две моды с $n=0$), а также дважды вырожденное движение в направлении, перпендикулярном оси нанотрубки (соответствует $n=1$). При «виртуальном» образовании ДУНТ из двух ОУНТ половина голдстоуновских степеней свободы должна пропасть.

Сначала рассмотрим, как проявляется исчезновение пары из четырех ГСС, которые при $k=0$ соответствуют вращениям и сдвигам вдоль оси исходных нанотрубок. Так как сдвигать и вращать ДУНТ вдоль ее оси можно только как целое, то из вышеупомянутых четырех ГСС исчезают две их линейные комбинации: относительное вращение и относительный сдвиг нанотрубок, которые приобретают частоты (8) за счет ван-дер-Ваальсовых взаимодействий. Дисперсионные кривые, соответствующие данным модам, в спектре ДУНТ обозначены индексами *R* и *S* (см. рис. 1, случай *n*=0).

Аналогичным образом, относительное движение двух нанотрубок, образующих ДУНТ, в перпендикулярном к их общей оси направлении (естественно, с сохранением общего центра тяжести) больше не является ГСС и приводит к появлению возвращающей силы, обусловленной межслоевыми взаимодействиями. Поэтому частота дисперсионной ветки (обозначенной индексом *X*), содержащей эту моду, не может быть нулевой при *k*=0. В итоге у ДУНТ остается только одна дисперсионная ветка с *n*=1 (обозначенная индексом *Q*), частота которой стремится к нулю пропорционально $k^2$.

Заметим, что при «виртуальном» образовании ДУНТ из пары ОУНТ наиболее сильно взаимодействуют друг с другом пары близкорасположенных веток с преимущественно радиальной поляризацией и одинаковым значением *n*. Именно такие деформации внутренней и внешней трубок могут наиболее сильно «мешать» друг другу. В случае *n*≥1 (смотрите случаи *n*=1 и *n*=2 в двух нижних строчках рис. 1) ветка, в которой пара ОУНТ движется примерно синфазно, остается практически на старом месте, а ветка, примерно соответствующая противофазному движению нанотрубок, приподнимается. При *n*=0 (см. верхнюю строчку рис. 1) заметно приподнимаются обе ветки. Таким образом, РДМ двух исходных ОУНТ (см. рис. 1, первая строка, самые высокочастотные ветки) становятся ДПМ модами ДУНТ.

Теперь перейдем к обсуждению влияния радиального взаимодействия ДУНТ с окружением на ее колебательный спектр. Для того чтобы сделать это влияние заметным, мы положили величину $C/\rho \approx 4000$ s$^{-2}$ и перестроили дисперсионные кривые для ДУНТ (22, 14)@(40, 1) (рис. 1, последний столбец). Наибольшему частотному сдвигу (в сторону больших частот) оказываются подвержены моды, имеющие преимущественно радиальную поляризацию. В частности, при *n*=0 увеличиваются частоты ДПМ ДУНТ, а при *n*=1 ДУНТ теряет еще две ГСС (соответствующие двукратно вырожденному движению ДУНТ в плоскости, перпендикулярной ее оси). Вследствие этого дисперсионная ветка *Q* поднимается вверх, занимая новое положение *Q'*. Таким образом, радиальный пиннинг приводит к весьма существенному уменьшению плотности фононных состояний в самой низкочастотной области спектра и должен заметно уменьшать сверхнизкотемпературную теплоемкость ДУНТ. В случае ОУНТ этот эффект также приводит к возникновению подобной щели в фононном спектре и, как следствие, к уменьшению сверхнизкотемпературной теплоемкости ОУНТ более чем на порядок [10]. Причина данного явления кроется в том, что именно самая низкочастотная дисперсионная ветвь (аналогичная ветви *Q*) вносит наибольший (до 90%) вклад в теплоемкость ОУНТ при сверхнизких температурах. Вопрос о влиянии радиального взаимодействия с окружением на теплоемкость ДУНТ будет рассмотрен более подробно в следующем разделе.

## 4. Низкотемпературная теплоемкость ДУНТ

Любая континуальная модель применима вплоть до тех температур, которые возбуждают лишь низкочастотные, описываемые моделью фононные моды. Следуя [10], мы считаем, что модами с частотами, для которых соотношение

$$\frac{h\nu}{k_B} \leq 9T \qquad (14)$$

не выполняется, практически не возбуждаются температурой. В неравенстве (14) $h$ – постоянная Планка, $k_B$ – постоянная Больцмана. Учтя, что самая низкочастотная мода ДУНТ (мода $Q$, рис. 1) на границе области применимости модели $k_{eff}=k_{lim}=7$ nm$^{-1}$ имеет частоту около 210 cm$^{-1}$, мы получаем, что континуальную модель можно применять для анализа тепловых свойств ДУНТ вплоть до температур порядка 35 K.

Как следует из второго раздела, все колебательные моды ДУНТ с ненулевыми $n$ и $k$ четырежды вырождены, поскольку определитель (6) квадратичен по $n$ и $k$. Кроме того, для каждого $n$ и $k$ возможно 6 низкочастотных колебательных мод, при возбуждении которых смещения соседних внутри слоя атомов углерода отличаются незначительно. Далее нам следует учесть, что на каждый фонон из каждой дисперсионной ветки в обратном пространстве приходится интервал $2\pi/L$. Наконец, перейдя в распределении Бозе [20] от суммирования к интегрированию, в низкотемпературном пределе мы записываем полную колебательную энергию ДУНТ в виде:

$$U = \frac{L}{\pi} \sum_{n=-n_{\lim}}^{n=n_{\lim}} \sum_{j=1}^{6} \int_{0}^{k_0(n)} \frac{h\nu_j(k,n)}{\exp\left(\frac{h\nu_j^{(1)}(k,n)}{k_B T}\right) - 1} dk, \qquad (15)$$

где индекс $j$ пробегает по шести дисперсионным веткам, $n_{lim}=8$ (моды с большим $n$ всегда имеют частоту, неудовлетворяющую (14)) и $k_0(n)$ – максимальная длина волнового вектора, который при данном $n$ еще удовлетворяет области применимости модели (14). Дифференцируя выражение (15) по температуре и численно интегрируя его по $k$, получаем зависимость теплоемкости ДУНТ от температуры. В качестве примера для расчетов мы выбрали несоразмерную ДУНТ (22, 14)@(40, 1), поскольку мы можем оценить величину как радиального, так и тангенциального межслоевого взаимодействия на основе экспериментальных данных [19] (см. начало раздела 3 настоящей работы). На рис. 2(а) приведены соответствующие результаты: кривая с квадратными маркерами соответствует удельной теплоемкости данной ДУНТ, кривая с треугольными маркерами – сумме удельных теплоемкостей отдельных ОУНТ (22, 14) и (40, 1). Результаты расчетов показывают, что в области применимости модели ($T$<35 K) сумма удельных теплоемкостей отдельных ОУНТ оказывается заметно больше, чем удельная теплоемкость соответствующей ДУНТ. Данный результат обусловлен межслоевыми ван-дер-Ваальсовыми взаимодействиями в ДУНТ, которые весьма существенно повышают частоты некоторых мод в ее низкочастотном спектре (см. раздел 3).

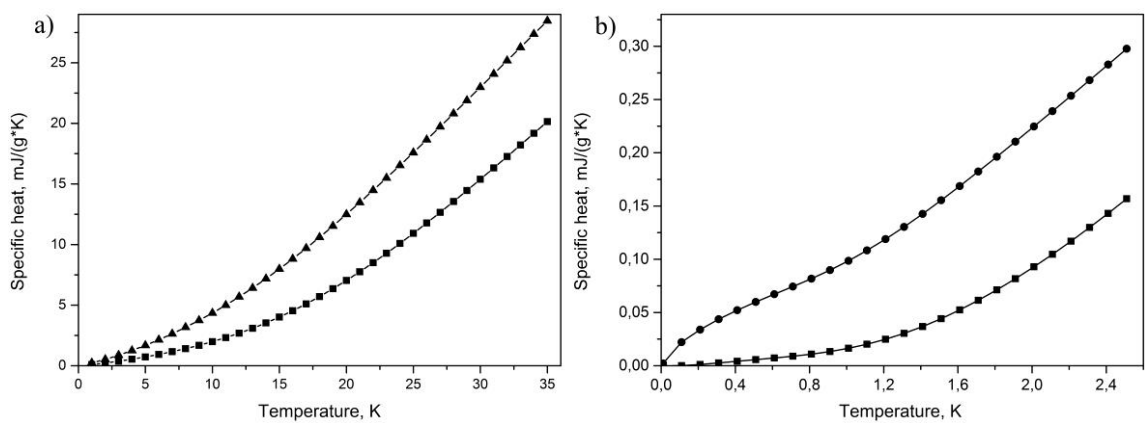

Рис. 2. (а) Результаты расчетов удельной теплоемкости ДУНТ (22, 14)@(40, 1): кривая с квадратными маркерами соответствует удельной теплоемкости данной ДУНТ при учете ван-дер-Ваальсовых взаимодействий между ее слоями; кривая с треугольными маркерами – сумме удельных теплоемкостей отдельных ОУНТ (22, 14) и (40, 1).
(b) Результаты расчетов удельной теплоемкости ДУНТ (22, 14)@(40, 1) при сверхнизких температурах без учета (кривая с круглыми маркерами) и с учетом (кривая с квадратными маркерами) взаимодействия между указанной ДУНТ и ее окружением. Величина $C/\rho$, определяющая радиальное взаимодействие ДУНТ с окружением, была принята равной 100 $s^{-2}$ (как в случае с ОУНТ в работе [10], где это значение было получено с помощью аппроксимации экспериментальных данных).

На рис. 2(б) представлены результаты расчетов удельной теплоемкости ДУНТ (22, 14)@(40, 1) в области сверхнизких температур: кривая с круглыми маркерами соответствует расчетам без учета взаимодействия между ДУНТ и ее окружением, кривая с квадратными маркерами – с его учетом. Оценить отношение $C/\rho$ из имеющихся на сегодняшний день экспериментальных данных не представляется возможным, поэтому мы приняли это отношение равным 100 $s^{-2}$. В [10] данная оценка была получена посредством анализа экспериментальных данных по теплоемкости пучков ОУНТ. При величине $C/\rho$ порядка 100 $s^{-2}$ расчет удельной теплоемкости ДУНТ показывает, что влияние окружения приводит к существенному (от 3 раз до более чем на порядок) падению теплоемкости ДУНТ только в области сверхнизких температур $T$<1.5 К. Данный эффект заметен в масштабе рис. 2 (б). При более высоких температурах $T$=1.5–35 К существенное (в 1.5–2.5 раза) уменьшение удельной теплоемкости ДУНТ (по сравнению с суммой теплоемкостей образующих ДУНТ нанотрубок) обусловлено, главным образом, межслоевыми ван-дер-Ваальсовыми взаимодействиями в ДУНТ, а не взаимодействием ДУНТ с окружением.

Отметим, что даже если пренебречь взаимодействием ДУНТ с окружением, то полностью корректно сравнить результаты наших расчетов с доступными на сегодняшний день экспериментальными данными [4, 5] не представляется возможным. Помимо сведений об удельной теплоемкости образца и средних диаметрах ДУНТ, содержащихся в нем, необходимо также располагать такими структурными характеристиками и данными спектроскопии КРС, которые позволили бы сопоставить внутренний и внешний диаметры присутствующих в образце ДУНТ с частотами ДПМ тех же самых ДУНТ. Без подобного рода данных достаточно точно определить величину межслоевого ван-дер-Ваальсова взаимодействия в ДУНТ (и, следовательно, корректно произвести расчет удельной теплоемкости того или иного образца) не представляется возможным. Подчеркнем еще раз, что оценка $G_r$ на основе соотношений (11) является очень грубой и может приводить к ошибкам в десятки процентов по сравнению со спектроскопической оценкой (на основе уравнения (7)) той же величины.

## 5. Заключение

В работе построена простейшая континуальная теория фононной динамики ДУНТ, взаимодействующей с окружением. На основе предлагаемого подхода проанализирована связь между низкочастотной частью колебательного спектра ДУНТ и спектрами составляющих ее одностенных нанотрубок. Полученные результаты использованы для расчета удельной теплоемкости ДУНТ в низкотемпературной области. Установлено, что удельная теплоемкость ДУНТ в области применимости предлагаемой континуальной модели ($T$<35 K) существенно (в 1.5–2.5 раза) ниже суммы удельных теплоемкостей составляющих ее ОУНТ. К такому уменьшению теплоемкости приводят ван-дер-Ваальсовы взаимодействия радиального и тангенциального типов между слоями ДУНТ. Если в двух невзаимодействующих ОУНТ существует всего шесть фононных веток, частоты которых стремятся к нулю вместе с волновым вектором, то в результате учета межслоевых взаимодействий число таких веток сокращается в два раза. В результате плотность фононных состояний в низкочастотном спектре ДУНТ существенно уменьшается, что и ведет к падению низкотемпературной теплоемкости.

В настоящее время теплоемкость индивидуальных нанотрубок все еще не может быть измерена. Поэтому в работе предложен феноменологический способ учета взаимодействия между индивидуальной ДУНТ и ее окружением, которое в наиболее значительной степени «препятствует» колебательным модам нанотрубки с преимущественно радиальной поляризацией. Данный эффект заметно (от 3 раз до более чем на порядок) уменьшает теплоемкость ДУНТ только в области сверхнизких температур ($T$<1.5 K). Представленная в данной работе теория может служить основой для дальнейшего изучения низкотемпературных тепловых характеристик как ДУНТ, так и других двустенных и многостенных тубулярных структур.

**Список литературы**
[1] C. Shen, A. H. Brozena, Yu. Wang. Nanoscale **3**, 503 (2011).
[2] Y. A. Kim, K.-S. Yang, H. Muramatsu, T. Hayashi, M. Endo, M. Terrones and M. S. Dresselhaus. Carbon Letters **15** (2), 77 (2014).
[3] G. Silva, A. W. Musumeci, A. P. Gomes, J.-W. Liu, E. R. Waclawik, G. A. George, R. L. Frost, M. A. Pimenta. Journal of Material Science **44**, 3498 (2009).
[4] W. Shi, Z. Wang, Q. Zhang, Y. Zheng, C. Ieong, M. He, R. Lortz, Y. Cai, N. Wang, T. Zhang, H. Zhang, Z. Tang, P. Sheng, H. Muramatsu, Y. A. Kim, M. Endo, P. T. Araujo, M. S. Dresselhaus. Scientific Reports **2**, 625 (2012).
[5] B. Xiang, C. B. Tsai, C. J. Lee, D. P. Yu, Y. Y. Chen. Solid State Communications **138**, 516 (2006).
[6] M. Damnjanović, I. Milošević, E. Dobardžić, T. Vuković, B. Nikolić. Physical Review B **69**, 153401 (2004).
[7] C. Li, T.-W. Chou. Materials Science and Engineering A **409**, 140 (2005).
[8] S. B. Rochal, V. L. Lorman, Yu. I. Yuzyuk. Physical Review B **88**, 235435 (2013).
[9] H. Suzuura and T. Ando. Physical Review B **65**, 235412 (2002).
[10] M. V. Avramenko, I. Yu. Golushko, A. E. Myasnikova, S. B. Rochal. Physica E **68**, 133 (2015).
[11] P. B. Canham. Journal of Theoretical Biology **26** (1), 61 (1970).
[12] W. Helfrich. Zeitschrift für Naturforschung C **98** (11), 693 (1973).
[13] S.V. Goupalov. Physical Review B **71**, 085420 (2005).
[14] F. Villalpando-Paez, H. Son, D. Nezich, Y. P. Hsieh, J. Kong, Y. A. Kim, D. Shimamoto, H. Muramatsu, T. Hayashi, M. Endo, M. Terrones, M. S. Dresselhaus. Nano Lett. **8**, 3879 (2008).
[15] T. Ch. Hirschmann, P. T. Araujo, H. Muramatsu, X. Zhang, K. Nielsch, Y. A. Kim, and M. S. Dresselhaus. ACS Nano **7**, 2381 (2013).
[16] M. Xia, S. Zhang, X. Zuo, E. Zhang, S. Zhao, J. Li, L. Zhang, Y. Liu, R. Liang. Physical Review B **70**, 205428 (2004).
[17] J. F. Nye. Physical Properties of Crystals: Their Representation by Tensors and Matrices. Oxford University Press (1985). 352 с.
[18] A. Bosak, M. Krisch, M. Mohr, J. Maultzsch, C. Thomsen. Physical Review B **75**, 153408 (2007).
[19] K. Liu, X. Hong, M. Wu, F. Xiao, W. Wang, X. Bai, J. W. Ager, S. Aloni, A. Zettl, E. Wang, F. Wang. Nature Communications **4**, 1375 (2013).
[20] Ch. Kittel. Introduction to Solid State Physics. Wiley, USA, (2005). 704 с.